\documentclass[aps,twocolumn,superscriptaddress]{revtex4}

\usepackage{graphicx}
\usepackage{amssymb,amsmath}
\usepackage[utf8]{inputenc}
\usepackage{bm}

\usepackage{psfrag}
\usepackage{yfonts}

\usepackage{version}

\usepackage{natbib}

\usepackage{longtable}

\newcommand{\eref}[1]{Eq.~(\ref{#1})}

\newcommand{\cref}[1]{chapter~\ref{#1}}

\newcommand{\Cref}[1]{Chapter~\ref{#1}}

\newlength{\mylenunit}
\begin{document}

\title{Vibronic Lineshapes of PTCDA Oligomers in Helium Nanodroplets}

\author{Jan Roden} 
\affiliation{Max-Planck-Institut f\"ur Physik komplexer Systeme, N\"othnitzer Str.~38, D-01187 Dresden, Germany}

\author{Alexander Eisfeld}
\affiliation{Max-Planck-Institut f\"ur Physik komplexer Systeme, N\"othnitzer Str.~38, D-01187 Dresden, Germany}

\author{Matthieu Dvo\v{r}\'{a}k}
\affiliation{Physikalisches Institut, Universit\"at Freiburg, Hermann-Herder-Str.~3, D-79104 Freiburg, Germany}

\author{Oliver B\"{u}nermann}
\affiliation{Institut f\"ur Physikalische Chemie, Universit\"at G\"ottingen, Tammannstr.~6, D-37077 G\"ottingen, Germany}
\affiliation{Physikalisches Institut, Universit\"at Freiburg, Hermann-Herder-Str.~3, D-79104 Freiburg, Germany}

\author{Frank Stienkemeier}
\affiliation{Physikalisches Institut, Universit\"at Freiburg, Hermann-Herder-Str.~3, D-79104 Freiburg, Germany}


\begin{abstract}
Oligomers of the organic semiconductor PTCDA are studied by means of helium nanodroplet isolation (HENDI) spectroscopy. 
In contrast to the monomer absorption spectrum, which exhibits clearly separated, very sharp absorption lines, it is found that the oligomer spectrum consists of three main peaks having an apparent width orders of magnitude larger than the width of the monomer lines. 
Using a simple theoretical model for the oligomer, in which a Frenkel exciton couples to internal vibrational modes of the monomers, these experimental findings are nicely reproduced. 
The three peaks present in the oligomer spectrum can already be obtained taking only one effective vibrational mode of the PTCDA molecule into account. 
The inclusion of  more vibrational modes leads to  quasi continuous spectra,
resembling the broad oligomer spectra.
\end{abstract}

\keywords{helium nanodroplets, PTCDA,
  Frenkel exciton, vibronic coupling}

\maketitle

\newcommand{\bra}[1]{\langle\,{#1}\, |}
\newcommand{\ket}[1]{|\,{#1}\,\rangle}
\newcommand{\braket}[2]{\mbox{$\langle\,{#1}\, | \,{#2}\,\rangle$}}
\newcommand{\ketbra}[2]{|\,{#1}\,\rangle\langle\,{#2}\,|}
\newcommand{\eps}{\mbox{\boldmath $\epsilon$}}

\newcommand{\lrb}[1]{\langle\, {#1}\,\rangle}
\newcommand{\Erw}[1]{\big[\!\!\big[{#1}\big]\!\!\big]}
\newcommand{\Real}{\mbox{Re}}

\newcommand{\V}{V}
\newcommand{\ElTransE}{\epsilon}
\newcommand{\HamTot}{H}
\newcommand{\transEnergy}{\epsilon}
\newcommand{\meanTrans}{{\ElTransE}}
\newcommand{\width}{\sigma}
\newcommand{\variance}{\sigma^2}
\newcommand{\transitEAgg}{\ElTransE_{\rm \tiny Agg}}
\newcommand{\ScalePar}{\sigma}
\newcommand{\StabIndex}{\alpha}
\newcommand{\Distrib}{S}
\newcommand{\FWHM}{\Delta}
\newcommand{\p}{\mathcal{P}}
\newcommand{\W}{W}
\newcommand{\Kern}{Q}

\newcommand{\LichtPol} {\hat{\mathcal{E}}}

\newcommand{ \dipOp} { \vec{D}}   

\newcommand{ \dip}{ \vec{d}}
\newcommand{\e}{\mbox{e}}

\newcommand{\om}{\omega}

\newcommand{\al}{\alpha}
\newcommand{\cm}{\ {\rm cm}^{-1}}
\newcommand{\K}{\ {\rm K}}

\section{Introduction}

Over the years there has been a lot of interest in the formation of finite molecular aggregates~\cite{WePe65_1894_,KoHaKa81_498_,BaCaMo02_149_,SaEs08_471_,SeWiRe09_13475_,GuZuCh09_154302_}. 
In the investigation of such aggregates optical spectroscopy plays an important role, since the aggregation manifests itself in (often drastic) changes of the absorption spectra~\cite{Sc38_1_,MoDaeDu95_6362_,Sp09_4267_,SpDaeOu00_8664_,EiBr06_376_,KiDa06_20363_,ReWaNe93_715_}. Usually in these studies the considered aggregates are embedded in a solvent at room temperature causing considerable additional broadening of the individual vibronic absorption lines~\cite{AkOezZh96_14390_,Kl81_251_,ReWi97_7977_}. Although such optical investigations already provide useful information e.g.\ on dimerisation constants, many details are not available due to the high temperatures and strong  interactions of the aggregates with the environment (solvent).

Using helium nanodroplet isolation spectroscopy it has become possible to
study aggregation in an environment (liquid helium) which interacts only
weakly with the chromophores. As shown in Refs.~\cite{WeSt03_125201_,WeSt04_1239_,WeSt05_1171_} this technique allows to obtain monomer spectra of organic molecules with  individually resolved vibronic lines. In this work we investigate absorption spectra of oligomers of the organic semiconductor PTCDA in helium nanodroplets. PTCDA (3,4,9,10-perylene-tetracarboxylic-dianhydride, C$_{24}$H$_{8}$O$_{6}$) is one of the most studied perylene derivatives. Its semiconducting properties and its ability to form highly structured crystalline films on different substrates~\cite{forrest_epitaxy}  makes this planar molecule interesting for applications~\cite{forrest_review}, such as organic light emitting diodes~\cite{grem1992,coe2002}, thin-film transistors~\cite{horowitz1998,dimitrakopoulos2002,murphy2007,organo_electro} or dye-sensitized solar cells~\cite{peumans2000,hoppe2004,hardin2009,organo_photovolt}. Most of the measurements on PTCDA concern thin films or crystals: highly ordered films of PTCDA have been grown and studied on poly- or single crystalline substrates like mica and gold~\cite{proehl2005}, silver~\cite{schneider2006}, graphene~\cite{huang2009}, aluminium, titanium, indium and tin~\cite{hirose1996}. Raman spectroscopy of epitaxial films~\cite{scholz_raman}, electroabsorption~\cite{haskal1995}, photoluminescence~\cite{yu_kobitski2003} and Ultraviolet Photoelectron Spectroscopy (UPS) measurements on gold, silver and copper substrates~\cite{Duhm2008} were also reported.

Isolated PTCDA molecules have already been studied in different liquid solvents~\cite{bulovic_DMSO}, in solid SiO$_2$ matrix~\cite{engel2006}, on optical nanofibers~\cite{stiebeiner2009} (sub-monolayer deposition) and with UPS in the gas phase~\cite{dori2006,sauther2009}. In all these cases there is a strong interaction of the PTCDA with an environment.

\begin{figure}
 \includegraphics[width=0.7\mylenunit]{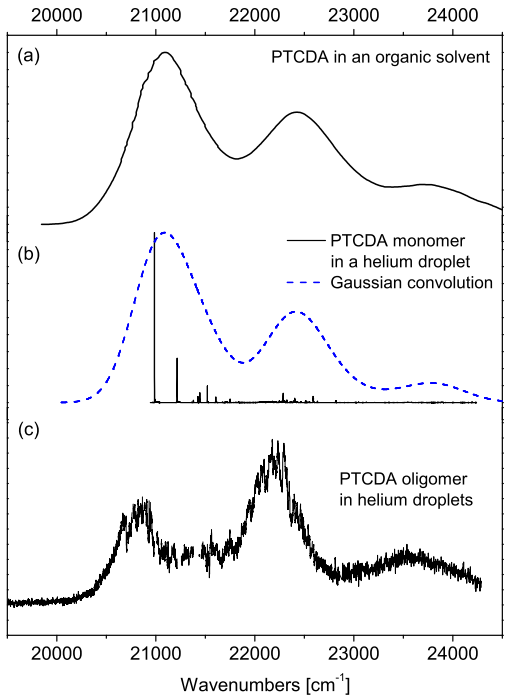}
   \caption{(a) Measured absorption spectrum of PTCDA monomers in an organic solvent (DMSO) at room temperature taken from Ref.~\cite{bulovic_DMSO}. 
For better comparison the shown spectrum is shifted by $1800\cm$ to higher wavenumbers.
(b) Measured PTCDA monomer fluorescence excitation spectrum in helium droplets~\cite{WeSt03_125201_} (solid line). 
The dashed curve is obtained by convoluting the measured spectrum with a Gaussian function having a width of $600\cm$ to facilitate comparison with the spectrum shown in (a).
(c) Measured PTCDA oligomer fluorescence excitation spectrum in helium nanodroplets}
  \label{fig:monomer_oligo_measured}
\end{figure}

In the present study PTCDA molecules are embedded in helium nanodroplets where they are only very weakly perturbed, which makes them ideal systems to compare with theoretical model calculations. The isolation of organic molecules in helium nanodroplets has been established as a versatile technique for studying absorption (see Ref.~\cite{Stienkemeier2001} and references therein) and emission~\cite{lehnig2003,lehnig2005} properties at high resolution. Spectroscopic measurements of single organic molecules and their derivatives with less than $1\cm$ resolution in the vibronic spectra have already been reported, providing detailed information on vibrational structures for example for perylene~\cite{carcabal2004}, PTCDA~\cite{WeSt03_125201_}, MePTCDI~\cite{WeSt05_1171_}, Phthalocyanine~\cite{lehnig2003} and some polyacenes~\cite{lehnig2005}. Figure~\ref{fig:monomer_oligo_measured} shows a comparison of an absorption spectrum of single PTCDA molecules isolated in helium droplets (panel (b)) with the corresponding spectrum taken in an organic solvent (panel (a)), demonstrating the huge difference in the resolution of the vibronic transitions.

Within the helium droplets several PTCDA molecules can aggregate to form oligomers. Such oligomers have been investigated by fluorescence absorption spectroscopy in order to study excitonic transitions~\cite{WeSt03_125201_}. It was found that in contrast to the narrow discrete monomer vibronic absorption lines (cf.~Fig.~\ref{fig:monomer_oligo_measured}(b)) the oligomer spectrum, although in helium nanodroplets, has broad peaks, more like oligomer spectra in ordinary solution or on a surface. So far, there has been no assignment of the individual peaks and despite of the clean low temperature conditions, there has been no theoretical concept explaining the outcome and widths of the spectra. In Fig.~\ref{fig:monomer_oligo_measured}(c) a corresponding PTCDA oligomer spectrum is shown. While in the previous studies the oligomer spectra where measured only over a limited region of excitation frequencies~\cite{WeSt03_125201_}, in the present work we could extend the covered  range up to 24300 cm$^{-1}$, revealing an additional peak at the high energy side.

In addition to extended experimental data, in this work we are able to reproduce the experimental findings using a simple Frenkel exciton model, that contains beside the dipole-dipole interaction between the monomers also the coupling of the electronic excitation to internal vibrations of the monomers. In particular, using this model, we will explain the appearance of the individual peaks of the oligomer spectrum and discuss the origin of their widths.

We emphasize that the aim of this work is not a detailed fit of the measured oligomer spectrum but to gain  basic insights into the underlying mechanisms. Therefore we adopt a minimalist model (for the PTCDA monomer as well as for the oligomer), being aware that further refinement could be necessary to obtain quantitative agreement between theory and experiment.

The paper is organized as follows: In Section~\ref{sec:experiment} the experimental setup is presented and details of the measured monomer and oligomer spectra are given. In Section~\ref{sec:model} we introduce the  model of the monomer and  the oligomer. Furthermore it is described how the absorption spectrum is calculated. In the following Section~\ref{sec:application} this theoretical model is applied to the calculation of the oligomer spectra. 
First the parameters needed to describe the PTCDA monomer are introduced. 
Since we cannot perform exact calculations of the oligomer spectra, including all the vibrational modes of PTCDA, in reasonable computing time, 
 the concept of effective modes (EM) is introduced and illustrated. Using these EMs we then demonstrate the dependence of the oligomer spectra on the dipole-dipole interaction and also on the number of implemented EMs. With the insight gained from these considerations we then model the measured absorption spectrum. In Section~\ref{sec:conclusion} we summarize our findings and conclude with an outlook.

\section{Experimental setup and findings}
\label{sec:experiment}

The experimental setup is similar to the one presented in an earlier
publication~\cite{WeSt05_1171_}. Helium droplets are produced by expanding
highly pressurized helium gas (purity $99.9999\%$, stagnation pressure 40 to
55 bars) into vacuum through a 10\,$\mu$m nozzle kept at a temperature of 17
to 19\,K by a closed-cycle refrigerator (Sumitomo RDK-408D). For these
measurements we choose expansion parameters which lead to helium droplets with
mean sizes of about 20\,000 atoms~\cite{ToeVil}. In a second vacuum chamber,
the helium droplets are doped with PTCDA molecules when flying through an oven
placed about 25\,cm downstream from the nozzle aperture. For a fixed helium
droplet size distribution, the number of PTCDA molecules picked-up by the
droplets depends on the density of PTCDA molecules in the oven given by its
temperature and the known vapor pressure. The mixture of different oligomer
sizes is given by the statistics of scattering events and follows a Poissonian
distribution~\cite{TiSt07_4748_}. At an oven temperature of
350\,$^{\circ}$C the droplets are preferentially doped with only single PTCDA
molecules (maximized monomer probability). For the oligomer spectrum presented
in Fig.~\ref{fig:monomer_oligo_measured}(c) the oven temperature was set to
390\,$^{\circ}$C. Poissonian statistics under these conditions
reveal~\cite{WeSt03_125201_} that approximately 8\% of the droplets are
undoped, 21\% of the droplets are doped with a single PTCDA molecule, 26\%
picked up two molecules, 21\% three, 13\% four and 6\% five molecules. In the
given spectrum (Fig.~\ref{fig:monomer_oligo_measured}(c)) the very sharp
monomer contributions have already been subtracted. Within the flight time to
the detectors all embedded oligomers are cooled by evaporating helium atoms to
the terminal droplet temperature of 380\ mK~\cite{Hartmann1999}. At these
temperatures, only vibrational ground states are populated. Shifts of
vibrational energies induced by the perturbation of the helium environment
have been determined to be less than 0.1\%~\cite{Callegari2001}. Shifts of electronic band origins of organic molecules
embedded in helium droplets have been observed in the range of 10~-- 100 wavenumbers~\cite{Stienkemeier2001}.

About 30\,cm further downstream, the doped droplet beam is crossed by a pulsed
laser beam for electronic excitation. The emitted Laser Induced Fluorescence
(LIF) light is collected and focused on a photomultiplier placed
perpendicularly to both the droplet beam and the laser beam. A second
photomultiplier aside the imaged fluorescence spot monitors the
stray light for efficient background suppression. The droplet beam is chopped
to allow for discrimination of PTCDA molecules coming directly from the
oven. In contrast to our former experiments we now used pulsed laser systems
having the capability to access a wider range of excitation frequencies. Two
laser systems have been used. A dye-laser pumped by the $3^{\rm rd}$ harmonic
(355\,nm) of a Nd:YAG pulsed laser (Edgewave IS-IIIE) covers the green-blue
range (500-370\,nm). A second dye-laser system pumped by the $2^{\rm nd}$ harmonic (532\,nm) of a pulsed Nd:YLF laser (Edgewave IS-IIE) provides output in the in the red-yellow region (590-500\,nm). Typically pulses are used of $\sim10$\,ns in widths at a repetition rate of 1\,kHz having a spectral resolution of around 3\,GHz.

\section{The theoretical model}
\label{sec:model}

In our model each PTCDA molecule is described by an electronic two level
system (according to the $S_0$ and $S_1$ state involved in the experiment). On
each monomer the electronic excitation couples linearly to a set of harmonic
vibrational modes. The oligomer is described by a Frenkel exciton Hamiltonian
with transition dipole-dipole interaction between the monomers~\cite{Fr30_198_,Fr31_17_}.  Other interactions (e.g.\ due to Van der Waals interaction,  overlap of electronic wavefunctions or interaction with the helium environment) are only considered via a global shift of the absorption spectra.

\subsection{Hamiltonian of a single monomer}
\label{sec:ham_op_mon}
First we consider a single monomer. Beside the electronic ground state  $\ket{\phi^g}$ we take one excited electronic state $\ket{\phi^e}$ into account.
In the Born-Oppenheimer approximation
we specialize to the particular case of harmonic vibrations of the
same frequency in upper and lower potential energy surface.
This allows for an efficient calculation of the oligomer spectrum (see
appendix).
Specifically, for each monomer we take $M$ vibrational modes into account.
The particular choice of the modes (or effective modes) will be discussed in Section~\ref{sec:application}.
For monomer $n$ the coordinate belonging to the $j$-th mode is denoted by $Q_{nj}$ and $P_{nj}$ denotes the corresponding momentum operator.

\begin{figure}
\includegraphics[width=0.35\mylenunit]{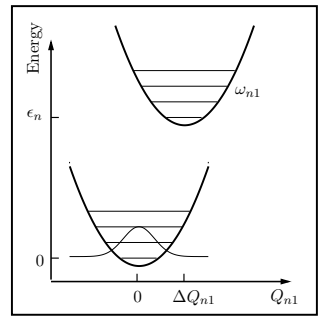}
\caption{\label{fig:BO_skizze}Sketch of the harmonic potentials in ground and
  excited electronic state of  monomer $n$ for one vibrational mode.}

\end{figure}

The vibrational frequency of mode $j$ is denoted by $\omega_{nj}$ and  the shift between the minima of the excited and ground state harmonic potential is denoted by
 $\Delta Q_{nj}$. The potential surfaces are sketched in Fig.~\ref{fig:BO_skizze}  for the case of one mode.
The vibrational Hamiltonian of monomer $n$ in the electronic ground state is then given by
\begin{equation}
\label{HamMonGround}
H^g_n=\frac{1}{2}\sum_{j=1}^M(P_{nj}^2+\om_{nj}^2 Q_{nj}^2)
\end{equation}
and for the excited electronic state
\begin{equation}
\label{HamMonExc}
H_n^e=\ElTransE_{n}+\frac{1}{2}\sum_{j=1}^M\big(P_{nj}^2+\om_{nj}^2
(\Kern_{nj}- \Delta Q_{nj})^2\big)
\end{equation}
where $\ElTransE_{n}$ denotes the energy difference between the minima of the
upper and lower potential energy surface.
We further introduce the dimensionless Huang-Rhys factor~\cite{MeOs95__}
\begin{equation}
  X_{nj}\equiv \frac{\om_{nj}}{2\hbar}(\Delta Q_{nj})^2,
\end{equation}
which measures the strength of the coupling of electronic excitation of monomer $n$ to vibrational mode $j$ of this monomer.

\subsection{Hamiltonian  of the aggregate}
\label{sec:H_nmer}

One of our basic assumptions is that the $N$ monomers forming the aggregate
have fixed positions and orientations w.r.t.\  each other.
That means that we neglect vibrational motion of the monomers w.r.t.\ each
other and also conformational changes that could appear upon electronic
excitation.

The electronic ground state of the aggregate is taken as the state
\begin{equation}
  \label{eq:ground_el_state}
  \ket{g_{\rm el}}=\ket{\phi^g_1}\cdots  \ket{\phi_N^g},
\end{equation}
in which all monomers are in their electronic ground state.
Since we consider  linear absorption it is sufficient to take only states into
account, in which at most one electronic excitation is on the aggregate,
i.e.\ states
\begin{equation}
  \ket{\pi_n}\equiv \ket{\phi^e_n}\prod_{m\neq n}^N\ket{\phi_m^g},
  \label{eq:pi_n}
\end{equation}
in which monomer $n$ is electronically excited and all other monomers are in their electronic ground state.

To keep the model plain,
we assume that in the aggregate the form of the monomeric potential energy surfaces is not changed.
Furthermore we will not consider overlap of electronic wavefunctions of different
monomers.
As the important interaction responsible for the lineshape of the
  oligomer we take the transition dipole-dipole interaction between the
monomers explicitly into account.
In addition we take energy shifts e.g.\ due to the changed (Van-der-Waals)
interaction between the monomers~\cite{Ei07_321_,AmVaGr00__} via an
overall shift of the calculated spectra into account.
Since one expects that the individual energy shifts depend on the number $N$ of monomers in the oligomer, we let also the included overall shift of the spectrum depend on $N$.

The  Hamiltonian in the electronic ground state of the aggregate is given by
\begin{equation}
\label{HamTotGround}
H^g=\left(\sum_{n=1}^N H_n^g \right)\ket{g_{\rm el}}\bra{g_{\rm el}}.
\end{equation}
We denote by $V_{nm}$ the transition dipole-dipole interaction between monomer
$n$ and $m$, which is assumed to be independent of nuclear coordinates.
 The Hamiltonian in the one-exciton space defined by the states
 Eq.~(\ref{eq:pi_n}) is then given by
\begin{equation}
\begin{split}
\label{HamTotExc}
H^e=&\sum_{n=1}^N \left(H_n^e  +\sum_{m\ne n}^N H_m^g
\right)\ket{\pi_n}\bra{\pi_n}\\
&+ \sum_{n,m=1}^N\V_{nm}\ket{\pi_n}\bra{\pi_m}.
\end{split}
\end{equation}

\subsection{Calculation of the aggregate absorption spectrum }

Since the temperature of the molecules inside the helium nanodroplets is below
$1\K$, which is more than two orders of magnitude less than the temperature corresponding to the energy quantum of the  vibrational mode with the lowest frequency ($230\cm$) of PTCDA, we assume that initially all vibrational modes of the PTCDA molecules are in their ground vibrational state.
Thus, as the initial state of the aggregate we take the state
\begin{equation}
 \ket{\Phi(t=0)}= \ket{g_{\rm el}}\ket{g_{\rm vib}},
\end{equation}
where all monomers are in their electronic ground state and
$\ket{g_{\rm vib}}=\prod_{n=1}^N\prod_{j=1}^M  \ket{\al_{nj}=0}$ with
$\ket{\al_{nj}=0}$ denoting the ground vibrational state of mode $j$ of monomer $n$.
We assume that the oligomers are isotropically oriented
and we denote by $\mu$ the magnitude of the transition dipoles of the
monomers.
The absorption strength for  light with frequency
 $\nu$  is then calculated from~\cite{La52_1752_,MaKue00__}
\begin{equation}
  A(\nu)= N |\mu|^2\ \Real\int_0^{\infty}dt\ \e^{i\nu t}\bra{\Psi_0}\e^{-iH^et/\hbar}\ket{\Psi_0}
  \label{eq:abs_spek_mit_tkorr_fkt}
\end{equation}
with
\begin{equation}
\label{PsiInitial}
 \ket{\Psi_0}=\frac{1}{\sqrt{N}}\sum_{n=1}^N\ket{\pi_n}\ket{g_{\rm vib}}.
\end{equation}
These formulas are valid if  all transition dipoles of the monomers are
identical, which for simplicity we have used in the following calculations.

Details of the numerical calculation can be found in the appendix.
At this point we only want to note that according to~\eref{eq:abs_spek_mit_tkorr_fkt} the spectrum is calculated by a Fourier transform of the correlation function $c(t)=\braket{\Psi_0}{\Psi(t)}$ with $\ket{\Psi(t)}=\e^{-iH^et/\hbar}\ket{\Psi_0}$.
Practically this propagation is carried out only to a certain finite time, which determines the resolution of the final spectrum.
To guarantee that $c(t)$ decays to zero until the end of the propagation we multiply it with a Gaussian window function of width $\sigma_{\tau}$, which results in a convolution of the stick spectrum with a Gaussian of the reciprocal width $\sigma=1/\sigma_{\tau}$.
For the spectra shown in the following  this width is indicated.

\section{Application of the formalism to PTCDA oligomers in He nanodroplets}
\label{sec:application}

\subsection{The monomer parameters}
To apply the  formalism of Section~\ref{sec:model} we need to know the frequencies
$\om_{nj}$ and the Huang-Rhys factors $X_{nj}$ of the various modes of
the PTCDA monomer.
However,  the numerical evaluation of \eref{eq:abs_spek_mit_tkorr_fkt} rapidly exceeds computer capabilities, even for
the dimer, if many modes that couple to the electronic excitation are
considered.
For the calculations we therefore group several modes $i$ with frequencies $\omega_i$ that lie in a certain frequency interval  together to form an effective mode (EM) $j$ with a new Huang-Rhys factor
\begin{equation}
  X_{j}^{\rm eff}=\sum_i X_{i},
\end{equation}
which is the sum of the Huang-Rhys factors $X_i$ of the original modes $i$.
This new EM $j$ has a frequency
\begin{equation}
  \omega_j^{\rm eff}= \frac{1}{X_j^{\rm eff}} \sum_{i}X_{i}\omega_{i}
\end{equation}
being the (Huang-Rhys factor weighted) average of the frequencies $\omega_{i}$ of the original modes $i$.
This procedure has also been applied e.g.\ in Ref.~\cite{GiSc09_115309_}.
However, one should  keep in mind that introducing the EMs is an approximation to the
original spectral density. Of course, the more modes are combined to
form an EM the more details of the original spectrum will be
missing. We will discuss this point in more detail later in this section.

Note also that the numerical effort for solving the {\it oligomer} problem not only depends on the
 number of EMs taken into account but also on the values of the Huang-Rhys factors.
This is because the stronger the coupling of the electronic excitation to a vibrational mode is, the more vibrational states of this mode have to be taken into account in our basis for the numerical calculation of the spectrum.
Due to the fact that the Huang-Rhys factors of the EMs are the sum of individual
Huang-Rhys (HR) factors of the original modes, the HR factors of the EMs will grow when reducing the number of EMs.
For larger HR factors one needs a larger basis, which limits the number $N$ of monomers of the oligomer that can be treated.

\begin{figure}[t]
  \includegraphics[width=0.8\mylenunit]{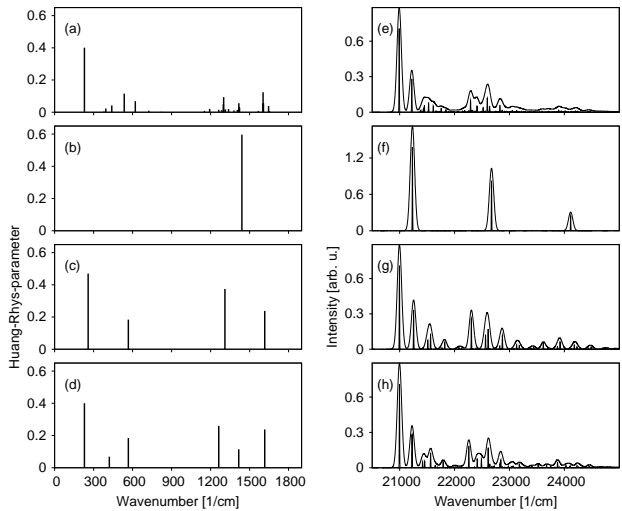}
  \caption{\label{fig:spec_dens}Left column: spectral density of vibrational modes. Right column: corresponding calculated monomer spectra. a) All 31 modes of Ref.~\cite{WeSt04_1239_}. b) One EM combining the modes in the frequency interval 1000-1800$\cm$ of (a). c) 4 EMs combining all modes of (a). d) 6 EMs combining all modes of (a).
The spectra in the right column are shown as sharp peaks and convoluted with a Gaussian of ${\rm FWHM}=100\cm$.}
\end{figure}

When constructing the EMs we take the frequencies and Huang-Rhys factors
extracted from the measured monomer spectrum
Fig.~\ref{fig:monomer_oligo_measured}b, that are given in Ref.~\cite{WeSt04_1239_},  as starting point.
Although it was noted in Ref.~\cite{WeSt04_1239_} that the peak intensities of the monomer spectrum do not
exactly follow a Poissonian, as expected for harmonic modes, for the purpose of the present paper these deviations are of no relevance. The frequencies $\om_j$ and corresponding Huang-Rhys factors $X_j$ for all 31 modes found in
Ref.~\cite{WeSt04_1239_} are shown in Fig.~\ref{fig:spec_dens}a and are listed in column~A of Tab.~\ref{tab:specdens}. Also shown  is the corresponding monomer absorption spectrum obtained from the harmonic model (see Fig.~\ref{fig:spec_dens}e). For better visibility in the monomer spectrum the sharp peaks are also convoluted with a narrow Gaussian.

An often adopted approach when considering oligomers in aqueous solution is to use only one EM describing the effect of the high energy vibrations~\cite{WiMo60_872_,Me63_154_,FuGo64_2280_,Wi65_161_,ScFi84_269_,EiBrSt05_134103_,SeMaEn06_354_,SeWiRe09_13475_}, corresponding to the progression of the broad maxima as seen in Fig.~\ref{fig:monomer_oligo_measured}(a).
In Fig.~\ref{fig:spec_dens}b we have illustrated this approach, combining the high energy modes of the interval 1000-1800$\cm$ of Fig.~\ref{fig:spec_dens}a, to obtain one EM. 
Also shown is  the
resulting monomer spectrum (see Fig.~\ref{fig:spec_dens}f).
Note the difference to the original spectrum
Fig.~\ref{fig:spec_dens}e.
We will use this model with only one EM to explain the three main peaks of the measured
oligomer spectrum.

\begin{table}
\begin{longtable}{c|c||c|c||c|c||c|c}
\multicolumn{2}{c||}{A}&\multicolumn{2}{c||}{B}&\multicolumn{2}{c||}{C}&\multicolumn{2}{c}{D}\\
\hline
$\omega$&$X$&$\omega$&$X$&$\omega$&$X$&$\omega$&$X$\\\hline\hline
229.5   & 0.401	& & &            &  &                 229.5 &  0.4 \\ \cline{7-8}
378.3   & 0.004 & & &            258& 0.47&                          &     \\
393.6	& 0.0229 & & &          & &                     421    & 0.07        \\
439.9   & 0.0421 & & &          & &                          &        \\ \cline{5-8}
535.3   & 0.115 & & &           567 &  0.18 &      567 & 0.18 \\
620.8   & 0.069 & & &           & &                           &        \\ \cline{5-8}
724.1   & 0.0097 & & &            &  &                   &        \\
823.2   & 0.0032 & & &            &  &                   &        \\ \cline{3-4}
1152.2  & 0.0085 & & &              & &                            &  \\
1192.2  & 0.0209 & & &          & &                               &  \\
1262.4  & 0.0155 & & &          & &                                &  \\
1284.8 	& 0.0121 & & &          & &                            1263  & 0.26  \\
1298.1 	& 0.0495 & & &          & &                              &  \\
1300.3 	& 0.0937 & & &          1310 & 0.37&                               &  \\
1314.3 	& 0.0172 & & &           & &                               &  \\
1338.1 	& 0.0192 & & &          & &                                 &  \\
1379.2 	& 0.0101 &1441 &0.6 &          & &                                 &  \\ \cline{7-8}
1405.6 	& 0.0137 & & &          & &                         &        \\
1413.2 	& 0.0167 & & &           & &                      1416 & 0.11 \\
1417.0 	& 0.056 & & &            & &                                  &  \\
1422.6 	& 0.0284 & & &           & &                                &  \\\cline{5-8}
1567.2 	& 0.0075 & & &          & &                      &              \\
1583.8 	& 0.002 & & &           & &                                &     \\
1603.5 	& 0.124 & & &           & &                      &  \\
1606.3 	& 0.0555 & & &         1616 &0.24 &                           1616      & 0.24 \\
1646.0 	& 0.0384 & & &          & &                                  &  \\
1651.3 	& 0.0076 & & &          & &                                  &  \\ \cline{3-4}
1919.3 	& 0.0007 & & &           & &                                  &  \\
1953.4 	& 0.0008 & & &          & &                                    &  \\
2060.6  & 0.0005 & & &          & &                             \\ 
2176.7  & 0.0008 & & &          & &                             \\ \hline
\end{longtable}
\addtocounter{table}{-1}
\caption{\label{tab:specdens} The four spectral densities used.
Shown are the values of the mode frequencies $\omega$ (in $\cm$) and the corresponding Huang-Rhys factors $X$.
Column A: Values of all 31 modes of Ref.~\cite{WeSt04_1239_} (shown also in Fig.~\ref{fig:spec_dens}a).
Column B: Values of the one EM (shown in Fig.~\ref{fig:spec_dens}b).
Column C: Values of the 4 EM (shown in Fig.~\ref{fig:spec_dens}c).
Column D: Values of the 6 EM (shown in Fig.~\ref{fig:spec_dens}d).
The horizontal lines indicate which modes of column A are comprised in the respective EM in the other columns.}
\end{table}

To obtain a more realistic description of a monomer one would like to take as many EMs  as possible.
The number of EMs that can be taken into account in our
numerical simulation strongly depends on the number of monomers within the  oligomer.
The spectrum of a  dimer with 6-7 modes can be calculated within one hour on a standard PC.
In a similar time, for a trimer we can only take 3 modes into account.
In the calculation of the spectra shown in the following we have used up to 6
modes for the dimer and 4 modes for the trimer (resulting in some increase in computer time).
As noted above the grouping into effective modes is quite arbitrary.
However, in the full spectral density Fig.~\ref{fig:spec_dens}a one notices
that several modes are densely packed together, like in the region around $1300\cm$.
Therefore it seems natural to put these modes together in one EM.
Similar for the modes around $500\cm$.
The remaining high energy modes around $1600\cm$ are also treated as one EM.

Particular sets of four/six effective modes  are shown in
Fig.~\ref{fig:spec_dens}c/Fig.~\ref{fig:spec_dens}d, which
 will be used in the following  calculations.
Of course the so  constructed sets of EMs are only one of many possible
choices, however  quite reasonable ones.
Note that the monomer absorption spectrum Fig.~\ref{fig:spec_dens}g for the case of four effective modes resembles the full spectrum Fig.~\ref{fig:spec_dens}e already quite well.

\subsection{Interpretation of the measured oligomer spectrum}
In this section we apply the foregoing model to explain the observed
features of the measured oligomer spectrum Fig.~\ref{fig:monomer_oligo_measured}c.

As described in Section~\ref{sec:experiment} the oligomer spectrum is expected to
consist mainly of dimer and trimer contributions.
Since the computational effort strongly increases with the oligomer size and since we do not aim to describe small
details of the measured spectra, we neglect in the following
the contribution of oligomers consisting of more than three monomers.
From the many possible geometrical arrangements we choose a particular simple
one. We consider a stack of  equidistantly spaced monomers with parallel
transition dipole moments.
Consequently we take the transition dipole-dipole interaction to be of the
form
$V_{nm}=V/|n-m|^3$ with a constant $V$.
As pointed out in the previous sections all other interactions are only
 taken into account via an overall $N$-dependent shift of the spectra.

\subsubsection{Model calculations}
\begin{figure}
\includegraphics[width=0.7\mylenunit]{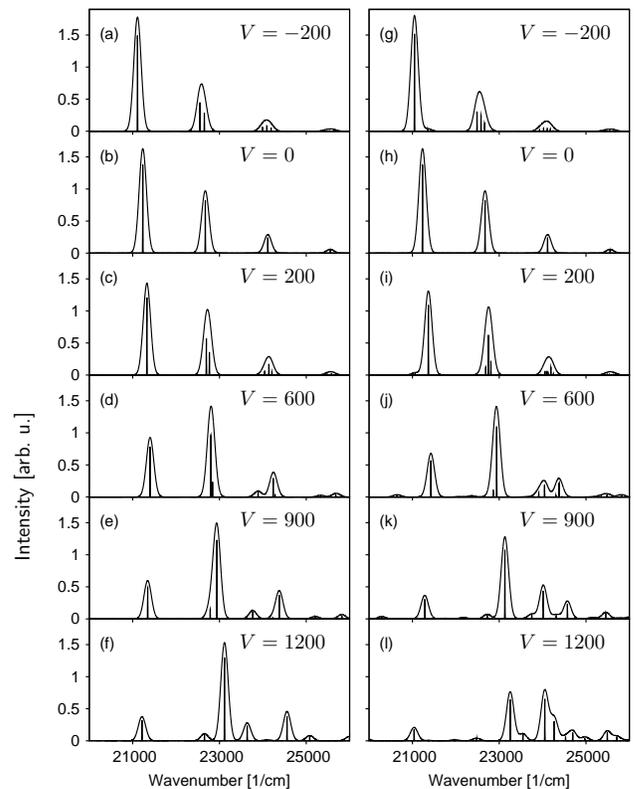}
\caption{\label{fig:dimer_1_mode}Dimer (left column) and trimer (right column) spectra for different interaction strength $V$ (in $\cm$).
Only the single vibrational mode of Fig.~\ref{fig:spec_dens}b is taken into account.
The spectra are shown as sharp peaks and convoluted with a Gaussian of ${\rm FWHM}=200\cm$ (to facilitate comparison of the overall intensities).}
\end{figure}

\begin{figure}
\includegraphics[width=0.7\mylenunit]{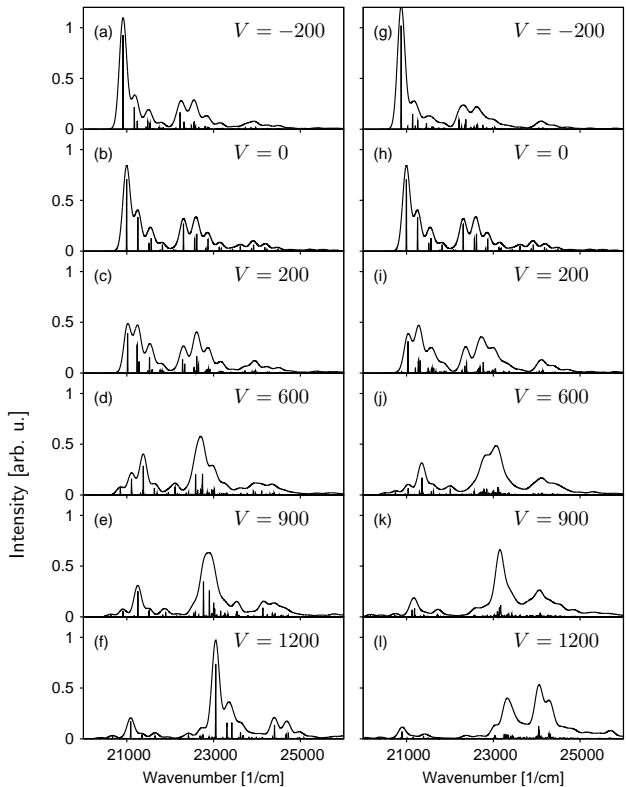}
\caption{\label{fig:dimer_5_mode}Same as Fig.~\ref{fig:dimer_1_mode} but with four modes (according to
Fig.~\ref{fig:spec_dens}c) taken into account.}
\end{figure}

To give an explanation of the appearance of the three peaks of the oligomer spectrum
we first consider only one effective high energy mode per monomer with $X$ and $\omega$ according to Fig.~\ref{fig:spec_dens}b
(i.e.\ $X=0.6$ and $\omega=1441\cm$).
Figure~\ref{fig:dimer_1_mode} shows dimer spectra (left column) and trimer spectra
(right column)  for various dipole-dipole interaction strength $V$ (indicated on the individual figures).
For small coupling $V=\pm 200\cm$ the spectra still look similar to the
  monomer spectrum ($V=0$). 
The main difference is a small splitting of the high energy peaks, which lifts the degeneracy of the energy levels present for $V=0$.
After convoluting with a Gaussian one notices only a small change in the
intensities of the resulting overall peaks upon  going from $V=0$ to $V=\pm 200\cm$.
However, for increasing $V$ this changes significantly.
For $V=600\cm$ we see in both cases, dimer and trimer, a peak structure (relative peak heights and distances between the peaks)
similar to that of the experimental spectrum Fig.~\ref{fig:monomer_oligo_measured}c.
Thus with one effective high energy mode one can already understand the
 the three peaks of the experimental spectrum.

If one takes now more  effective modes into
account one finds that a pronounced  ``broadening'' of the peaks appears. This
is demonstrated in
Fig.~\ref{fig:dimer_5_mode} where dimer and trimer spectra are shown  for
the same values of $V$ as in Fig.~\ref{fig:dimer_1_mode}, however, now including four
effective modes per monomer. The used spectral density is that shown in
Fig.~\ref{fig:spec_dens}c; the values for $X_j$ and $\omega_j$ are given also
in column~C of Tab.~\ref{tab:specdens}.
Note that now in certain energy regions (e.g.\ around $23000\cm$ for $V=600\cm$) many peaks are closely
spaced, forming a quasi continuum.
This effect becomes  especially pronounced in the high energy region of the spectra.
The appearance of the many densely packed peaks leads to an effective
broadening (but no significant change in position) of the three main peaks compared to
the case when only one mode is taken into account.
Thus to understand the broadening of the aggregate spectrum it is of great importance
to take as many internal modes as possible into account (this probably holds
also true for spectra of PTCDA in other solvents, however there interaction with the solvent
is expected to play an equally important role in broadening the spectra).

\subsubsection{Comparison to experiment}

\begin{figure}
\vspace{-1cm}
\includegraphics[width=0.7\mylenunit]{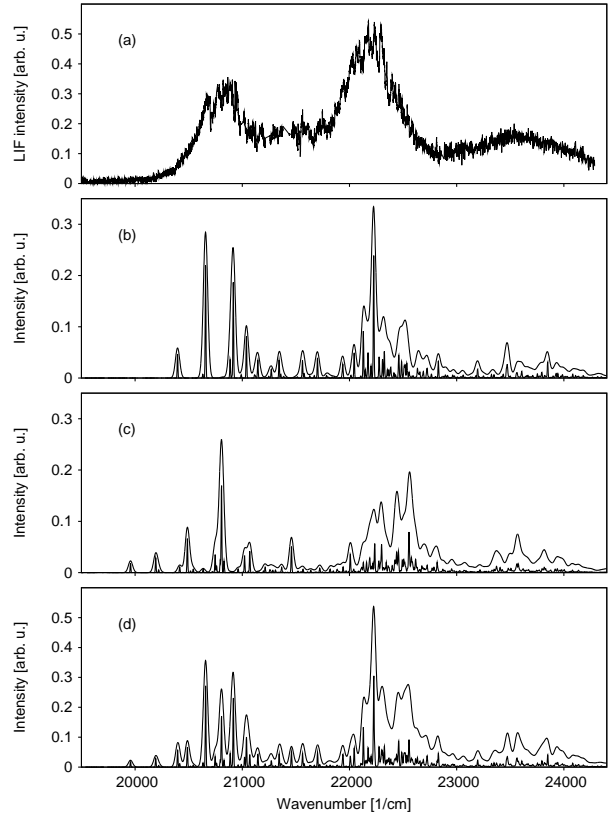}
\caption{\label{fig:compare}
 a) Measured oligomer spectrum. b-d) Calculated oligomer spectra for $V=600\cm$.
b) Calculated dimer spectrum for the spectral density of Fig.~\ref{fig:spec_dens}d, red-shifted by $450\cm$.
c) Calculated trimer spectrum for the spectral density of Fig.~\ref{fig:spec_dens}c, red-shifted by $550\cm$.
d) Sum of dimer spectrum (b) and trimer spectrum (c), where the two spectra are scaled according to the ratio $26/21$ of dimer and trimer contribution to the experimental oligomer spectrum (a).
The calculated spectra are shown as sharp peaks and in addition convoluted with a Gaussian of ${\rm FWHM}=50\cm$.}
\end{figure}

A direct comparison between calculated spectra and the experimental one is
given in Fig.~\ref{fig:compare}.
Here we have chosen $V=600\cm$, for which the relative heights and
distances of the three main peaks in the  measured  spectrum (shown in Fig.~\ref{fig:compare}a) are reproduced quite well.
The  dimer and trimer spectra (calculated including 6 EMs according to Fig.~\ref{fig:spec_dens}d for the dimer and 4 EMs according to Fig.~\ref{fig:spec_dens}c for the trimer) are then shifted to the red as a whole so that the peak positions coincide with the measured spectrum.
We have used a slightly larger red-shift for the trimer than for the dimer ($\Delta=550\cm$ and $\Delta=450\cm$ respectively).
As discussed in Section~\ref{sec:model} such shifts are expected to be e.g.\ due to Van der Waals
interactions.
The so obtained dimer spectrum  is shown Fig.~\ref{fig:compare}b, the trimer spectrum in
Fig.~\ref{fig:compare}c.
Finally, in Figure.~\ref{fig:compare}d a weighted sum of the dimer spectrum Fig.~\ref{fig:compare}b and trimer spectrum Fig.~\ref{fig:compare}c is shown.
The relative weights are 26/21 for dimer/trimer, taken to be consistent with the
ratio of droplets containing two and three PTCDA molecules.
To facilitate comparison with the measured oligomer spectrum we have convoluted the sharp peaks of the calculated oligomer spectrum with a narrow Gaussian of ${\rm FWHM}=50\cm$, much narrower than the apparent width of the three main peaks of the measured spectrum.
Clearly, this artificial broadening has no physical reason.
However, it facilitates the estimation of the overall intensity of the many sharp peaks.

The foregoing considerations demonstrate that the basic structure (relative heights,  positions and widths of the apparent peaks) of the
measured oligomer spectrum can be explained using a Frenkel exciton model,
where intra-monomer vibrations are taken into account.

In this work we have regarded the monomers to be in a fixed geometry.
However, upon electronic excitation inter-monomer motion could be induced.
The importance of such motion has been demonstrated in Ref.~\cite{FiSeEn08_12858_}.
This will lead to an additional broadening of the individual vibrational lines due to some low frequency motion (vibration/torsion/dissociation) of the monomers w.r.t.\ each other.

\section{Conclusions}
\label{sec:conclusion}

In this work we have presented a new extended LIF spectrum of PTCDA oligomers in helium nanodroplets. We have used a Frenkel exciton model including also intra-molecular vibrations to interpret the measured spectrum. Inclusion of one effective high-energy vibrational mode already allows to understand the appearance of the main features in the experimental data. Including more modes can explain at least part of the widths of the main peaks found in the experiment: While in the high energy region (above $21500\cm$) of the theoretical spectrum there is an already sufficiently large level density to explain the quasi-continuous measured spectrum, below  $21500\cm$ the density of levels is still too low to match the experimental findings. Since we have restricted  our calculations to dimers and trimers only, the inclusion of larger oligomers (which amount to roughly 1/4 of the measured oligomer spectrum) would give additional level density in this energy region.

The used simple model apparently catches all the main characteristics of the spectrum despite of severe approximations like e.g.\ fixed geometry and in particular not allowing charge transfer effects. In order to obtain a more detailed view of the type of oligomers present in the measured spectra these additional effects should be taken into account. However, since the experimental data are not oligomer size selected, at the present stage it seems not to be needful to introduce a more sophisticated model risking over-interpreting the experimental data.  Using hole-burning techniques it might be possible to obtain the individual contributions of dimers, trimers and higher aggregates and also to distinguish between different geometrical arrangements. Then, sum rules~\cite{Ei07_321_} might be helpful to obtain information on the conformation of the different oligomers species.

Nevertheless, the present study already reveals basic effects that will appear in the oligomer spectra. In particular, it demonstrates that it is important to include as many internal vibrations as possible in the model calculations. On the experimental side, vibrationally resolved spectra, as can be obtained by helium isolation spectroscopy, appear to be instrumental as input for modeling and understanding the absorption of such organic complexes.

\appendix
\section{Details of the numerical calculation}
\subsection{Representation of the Hamiltonian}
\label{sec:num_details}
We express the aggregate Hamiltonian using  creation and annihilation operators for the vibrational modes.
For the nuclear Hamiltonians  of monomer $n$ (see \eref{HamMonGround} and
\eref{HamMonExc}) one can write
\begin{equation}
  H_n^g=\sum^M_{j=1}\hbar\om_{nj} a_{nj}^{\dagger} a_{nj}
\label{eq:ham_g_mon}
\end{equation}
and
\begin{equation}
\begin{split}
  H_n^e=
   \tilde{\eps}_n+\sum^M_{j=1}\hbar\om_{nj}\left(a^{\dagger}_{nj} a_{nj}-\sqrt{X_{nj}}(a_{nj}^{\dagger}+a_{nj})\right),
\end{split}
\label{eq:ham_e_vib_mit_a_e_j}
\end{equation}
where we have incorporated an energy $\sum^M_{j=1}\hbar\om_{nj}X_{nj}$ into $\tilde{\eps}_n$.

The electronic ground state Hamiltonian of the aggregate \eref{HamTotGround} is then given by
\begin{equation}
  \label{eq:Agg_ground}
  H^g=\left(\sum_{n=1}^N \sum^M_{j=1}\hbar\om_{nj} a_{nj}^{\dagger} a_{nj} \right)\ketbra{g_{\rm el}}{g_{\rm el}}
\end{equation}
and for the Hamiltonian \eref{HamTotExc} in the one-exciton basis  one has
\begin{equation}
  H^e=H_{\rm el}+H_{\rm int}+H_{\rm vib}
  \label{eq:ham_el_ham_int_ham_vib}
\end{equation}
with a purely electronic part
\begin{equation}
  H_{\rm el}\equiv \sum_{n,m=1}^N \left(\tilde{\eps}_n\delta_{nm}+V_{nm}\right)\ketbra{\pi_n}{\pi_m}
  \label{eq:ham_el}
\end{equation}
and a vibrational part  given by
\begin{equation}
  H_{\rm vib}\equiv \sum^N_{n=1}\sum^M_{j=1}\hbar\om_{nj} a_{nj}^{\dagger} a_{nj}.
  \label{eq:ham_vib}
\end{equation}
The coupling between the electronic excitation and the vibrations is
contained in
\begin{equation}
  H_{\rm int}\equiv -\sum_{n=1}^N\ketbra{\pi_n}{\pi_n}\sum_{j=1}^M \hbar\om_{nj}\sqrt{X_{nj}}(a_{nj}^{\dagger}+a_{nj}).
  \label{eq:ham_int}
\end{equation}

\subsection{Calculation of the autocorrelation function}

To calculate the correlation function  $c(t)=\bra{\Psi_0}\e^{-iH^e t/\hbar}\ket{\Psi_0}$ , needed in
\eref{eq:abs_spek_mit_tkorr_fkt} to obtain the absorption spectrum, we numerically solve the time-dependent Schr\"odinger equation
\begin{equation}
  i\hbar\partial_t\ket{\Psi(t)}=H^e\ket{\Psi(t)}
  \label{eq:tdse}
\end{equation}
with initial state $\ket{\Psi(t=0)}=\ket{\Psi_0}$ given by Eq.~(\ref{PsiInitial}).
To this end we expand the full state $\ket{\Psi(t)}$ in the basis given by the states
\begin{equation}
  \ket{\theta_n^{\{\al\}}}\equiv \ket{\pi_n}\ket{\{\al\}}
  \label{eq:theta_n_al_basis}
\end{equation}
where the vibrational states $\ket{\{\al\}}$ are defined by
\begin{equation}
  \ket{\{\al\}}\equiv \prod_{n=1}^N\prod_{j=1}^M \ket{\al_{nj}}
  \label{eq:al_basis}
\end{equation}
with the states $\ket{\al_{nj}}$ being eigenstates of $a_{nj}^{\dagger} a_{nj}$, i.e.\
$
  a_{nj}^{\dagger} a_{nj}\ket{\al_{nj}} = \al_{nj}\ket{\al_{nj}}
$
with $\al_{nj}=0,1,\dots$, and the multi-index $\{\al\}$ contains all $\al_{nj}$ for all $n$ and $j$.
In this basis the matrix elements of the Hamiltonian $H^e$  are given by
\begin{equation}
\label{eq:agg_ham_matr_elem_theta_n_al_basis}
\begin{split}
  \bra{\theta_n^{\{\al\}}}H&\ket{\theta_m^{\{\beta\}}}  = \delta_{nm}\delta_{\{\al\}\{\beta\}}\left(\tilde{\eps}_n+ \sum^N_{k=1}\sum^M_{j=1}\hbar\om_{kj}\al_{kj}   \right)  \\
  &  - \
  \delta_{nm}\sum_{j=1}^M\frac{\delta_{\{\al\}\{\beta\}}}{\delta_{\al_{nj},\beta_{nj}}}\hbar\om_{nj}\sqrt{X_{nj}}\\
 &\times\left(\sqrt{\beta_{nj}+1}\ \delta_{\al_{nj},\beta_{nj}+1}+\sqrt{\beta_{nj}}\ \delta_{\al_{nj},\beta_{nj}-1}\right)\\
  &  + \ V_{nm}\delta_{\{\al\}\{\beta\}},
\end{split}
\end{equation}
where we have used the abbreviation
\begin{equation}
  \delta_{\{\al\}\{\beta\}}\equiv \prod_{n=1}^N\prod_{j=1}^M\delta_{\al_{nj},\beta_{nj}}.
\end{equation}
From \eref{eq:agg_ham_matr_elem_theta_n_al_basis} one sees that in the chosen
basis the Hamiltonian becomes very sparse, because of the factor $\delta_{\{\al\}\{\beta\}}$ contained in each term.
Finally, the
 time-dependent Schr\"odinger equation Eq.~(\ref{eq:tdse}) in this basis is given by
\begin{align}
\begin{split}
  &\braket{\theta_n^{\{\al\}}}{\dot{\Psi}(t)} =  -\frac{i}{\hbar}\left(\tilde{\eps}_n+ \sum^N_{k=1}\sum^M_{j=1}\hbar\om_{kj}\al_{kj}\right)\braket{\theta_n^{\{\al\}}}{\Psi(t)}  \nonumber\label{eq:tdse_in_theta_n_al_basis} \\
  &+\frac{i}{\hbar}\sum_{j=1}^M\hbar\om_{nj}\sqrt{X_{nj}}\sqrt{\al_{nj}}\ \braket{\theta_n^{\{\al_{11},\cdots ,\al_{nj}-1,\cdots\}}}{\Psi(t)}\\
& +\frac{i}{\hbar}\sum_{j=1}^M\hbar\om_{nj}\sqrt{X_{nj}}\sqrt{\al_{nj}+1}\ \braket{\theta_n^{\{\al_{11},\cdots ,\al_{nj}+1,\cdots\}}}{\Psi(t)}   \\
  &-\frac{i}{\hbar}\sum_{m\neq n}V_{nm}\braket{\theta_m^{\{\alpha\}}}{\Psi(t)}.
\end{split}
\end{align}

For the numerical calculation we take only certain relevant states of the (infinite) basis, defined in \eref{eq:theta_n_al_basis}, into account.
These states are chosen in the following way.
First we restrict the maximum number each $\alpha_{nj}$ can reach.
In the state $\ket{\pi_n}$, in which monomer $n$ is electronically excited, we take this number to be $Z_{nj}^e$, so that $\alpha_{nj}\leq Z_{nj}^e$.
In a state $\ket{\pi_m}$ with $n\neq m$, in which monomer $n$ is in its electronic ground state, we choose another maximum $Z_{nj}^g$, i.e.\ $\alpha_{nj}\leq Z_{nj}^g$.
Furthermore it has appeared to be advantageous to restrict the maximal number of vibrational states (and therewith the maximal vibrational energy) taken into account for a certain $j$.
We do this by setting a maximum $Z_j$ such that $\sum_{n=1}^N\al_{nj}\leq Z_j$ for the $j$-th mode.
The convergence of the calculated spectra is then tested by increasing one of the values $Z_{nj}^g$, $Z_{nj}^e$ or $Z_j$ at a time.

For example in the calculation of the trimer spectrum shown in Fig.~\ref{fig:compare}c we took $Z_1^g=2$, $Z_2^g=1$, $Z_3^g=2$ and $Z_4^g=1$ for the four used EMs given in column~C of Table~\ref{tab:specdens} (same values for all three monomers; the order of these values is the same as that of the respective mode frequencies in Table~\ref{tab:specdens}) and $Z^e=11$ for all modes and all monomers.
Additionally in this calculation we made the restriction $Z_1=3$, $Z_2=2$, $Z_3=3$ and $Z_4=2$ for the four modes.

For the calculation of the autocorrelation function we use the method described in Ref.~\cite{En92_76_} to increase the efficiency.

\begin{acknowledgments}
We thank John S.\ Briggs for organizing the Advanced Study Group "Quantum Aggregates" at the MPIPKS Dresden, that stimulated this collaboration.
\end{acknowledgments}

\vspace{0cm}

\end{document}